\documentclass[
reprint,
dcolumn,
bm,
nofootinbib,
notitlepage,
superscriptaddress]{revtex4-2}
\usepackage{graphicx}
\usepackage{xcolor}
\usepackage[latin1]{inputenc}
\usepackage[T1]{fontenc}
\usepackage{amsmath}
\usepackage{amssymb}
\usepackage{braket}
\usepackage{mathrsfs}
\usepackage{hyperref}
\usepackage{ulem}
\usepackage{upgreek}
\usepackage{pstricks-add}
\usepackage{natbib} 
\usepackage{listings}
\usepackage{tikz}
\usepackage{tcolorbox}
\usetikzlibrary{mindmap}

\definecolor{lightblue}{rgb}{0.145,0.6666,1}

\begin{document}
\title{Nucleophilic substitution at silicon under vibrational strong coupling: Refined insights from a high-level \textit{ab initio} perspective}

\author{Niels-Ole Frerick}
\affiliation{Institut f\"ur Chemie, Humboldt-Universit\"at zu Berlin, Brook-Taylor-Stra\ss{}e 2, D-12489, Berlin, Germany}

\author{Michael Roemelt}
\affiliation{Institut f\"ur Chemie, Humboldt-Universit\"at zu Berlin, Brook-Taylor-Stra\ss{}e 2, D-12489, Berlin, Germany}

\author{Eric W. Fischer}
\email{ericwfischer.sci@posteo.de}
\affiliation{Institut f\"ur Chemie, Humboldt-Universit\"at zu Berlin, Brook-Taylor-Stra\ss{}e 2, D-12489, Berlin, Germany}

\date{\today}

\let\newpage\relax

\begin{abstract}
We study the bimolecular nucleophilic substitution (S$_\mathrm{N}$2) reaction of 1-phenyl-2-trimethylsilylacetylene (PTA) under vibrational strong coupling (VSC) from the perspective of high-level \textit{ab initio} quantum and polaritonic chemistry. Specifically, we address conflicting mechanistic proposals, cavity-induced electronic corrections under VSC and the relevance of a previously debated Si-C-stretching motion of PTA for vibrational polariton formation. We first provide computational evidence for a two-step mechanism based on density functional theory and high-level coupled cluster results, identify new encounter and products complexes and illustrate the relevance of diffuse basis functions for a qualitatively correct description of anionic reactive systems. We subsequently show that cavity-induced dipole fluctuation corrections of electronic energies can be significant on the level of cavity Born-Oppenheimer coupled cluster theory and discuss their qualitative impact on the proposed two-step mechanism taking into account cavity-induced molecular reorientation. We finally show that the Si-C-stretching contribution to the experimentally relevant double-peak feature of PTA exhibits a dominant dipole character, which renders it central for linear IR response and vibrational polariton formation despite the presence of CH$_3$-rocking contributions. The dipole character along the cleaving Si-C-bond is eventually shown to rationalize Rabi splittings throughout the proposed two-step mechanism. Our work refines the microscopic perspective on the S$_\mathrm{N}$2 reaction of PTA under VSC and highlights recent developments in \textit{ab initio} polaritonic chemistry for the VSC regime. 
\end{abstract}

\let\newpage\relax
\maketitle
\newpage

\section{Introduction}
\label{sec.intro}
The experimental realization of strong light-matter coupling\cite{ebbesen2016} between molecular vibrations and confined field modes of an optical Fabry-P\'erot cavity led to the infrared (IR) spectroscopic characterization of vibrational polaritons in molecules\cite{shalabney2015,george2015,long2015}. In a series of seminal experimental studies, the vibrational strong coupling (VSC) regime was observed to intriguingly alter ground state chemistry\cite{thomas2016,thomas2019pta,thomas2019,xiang2020,ahn2023,patrahau2024,yin2025}, which ultimately culminated in the emerging field of vibro-polaritonic chemistry\cite{hirai2020,nagarajan2021,dunkelberger2022}. 

A paradigmatic example of vibro-polaritonic chemisty is the bimolecular nuclear substitution (S$_\mathrm{N}$2) reaction of 1-phenyl-2-trimethylsilylacetylene (PTA) with a fluoride anion in methanol as reported by Ebbesen and coworkers.\cite{thomas2016} Here, resonant VSC of an infrared cavity with a vibrational mode of PTA carrying Si-C-stretching character was experimentally observed to significantly reduce the rate of the S$_\mathrm{N}$2 reaction. In a subsequent experimental study, a dependence of related thermodynamic parameters on VSC was reported and a two-step mechanistic hypothesis including a stable pentacoordinate intermediate was proposed.\cite{thomas2019pta}

In a first computational study, Climent and Feist addressed both the microscopic S$_\mathrm{N}$2 reaction mechanism and the nature of vibrational modes in PTA.\cite{climent2020} The authors reported on a two-step mechanistic hypothesis with an intermediate pentacoordinate transition complex based on density functional theory (DFT) calculations in agreement with the experimental proposal. In addition, the experimentally relevant vibrational double-peak feature of PTA around $860\,\mathrm{cm}^{-1}$ was controversially assigned to be mainly of CH$_3$-rocking character with minor Si-C-stretching contributions.\cite{climent2020,thomas2020c1,climent2021c2}

Subsequently, Sch\"afer \textit{et al.}\cite{schaefer2022} presented an alternative single-step mechanistic hypothesis based on DFT methodology contradicting previous proposals\cite{thomas2019pta,climent2020}. The authors analysed the vibrational problem via semiclassical techniques with a strong focus on the resonance effect in the VSC regime and also reported on a dominant CH$_3$-rocking mode contribution for the spectroscopically relevant region of PTA. In a follow-up publication\cite{schaefer2024}, those results were partially revised from the perspective of machine-learning assisted \textit{ab initio} molecular dynamics and it was additionally suggested that previously neglected cavity-induced modifications of the electronic subsystem could actually be relevant. 

In this work, we aim at closing the conceptual gap between existing theoretical studies. To this end, we present a quantum-chemically inspired refined theoretical analysis of the PTA-Fluoride system from the perspective of high-level \textit{ab initio} methods capable of accurately describing the electronic subsystem also under VSC conditions. In a first step, we propose a two-step mechanistic hypothesis on a coupled cluster level of theory supporting result of Refs.\cite{thomas2019pta,climent2020}, which we augment by previously unidentified encounter and product complexes relevant for chemistry in solution. In this context, we show that adequately large basis sets with diffuse components are crucial for a qualitatively correct quantum chemical description of anionic reactive species in the S$_\mathrm{N}$2 reaction.  

We then address cavity-induced electronic corrections based on recent developments of \textit{ab initio} polaritonic chemistry in the cavity Born-Oppenheimer (CBO) framework\cite{flick2017,flick2017cbo,fischer2023,schnappinger2023,angelico2023,fischer2024ele,fischer2025}. Specifically, we show that cavity-induced dipole fluctuation corrections can be significant on a CBO coupled cluster level of theory in line with previous theoretical results\cite{fischer2024ele,fischer2025} and discuss their qualitative impact on the S$_\mathrm{N}$2 reaction under VSC . 

We finally reconsider the vibrational double-peak feature in the linear IR spectrum of PTA with a focus on optical properties of modes with a Si-C-stretching contribution. We find the Si-C-stretching contribution to the double-peak feature to be characterized by a strong dipole component along the Si-C bond, which is central for both its strong IR response and dominant contribution to vibrational polariton formation despite the presence of CH$_3$-rocking contributions discussed previously\cite{climent2020,schaefer2022}. Based on recently-developed linear response techniques\cite{fischer2024vib}, we furthermore show that dipole properties of normal modes are correlated with the Rabi splitting of reactive species throughout the S$_\mathrm{N}$2 reaction, which characterizes them as a conclusive indicator for a vibrational mode's relevance in VSC.

\section{Results and Discussion}
\label{sec.results}

\subsection{Refined Mechanistic Hypothesis}
\label{sec.mechanist_hypo}
\begin{figure*}[hbt!]
\begin{center}
\includegraphics[scale=1.0]{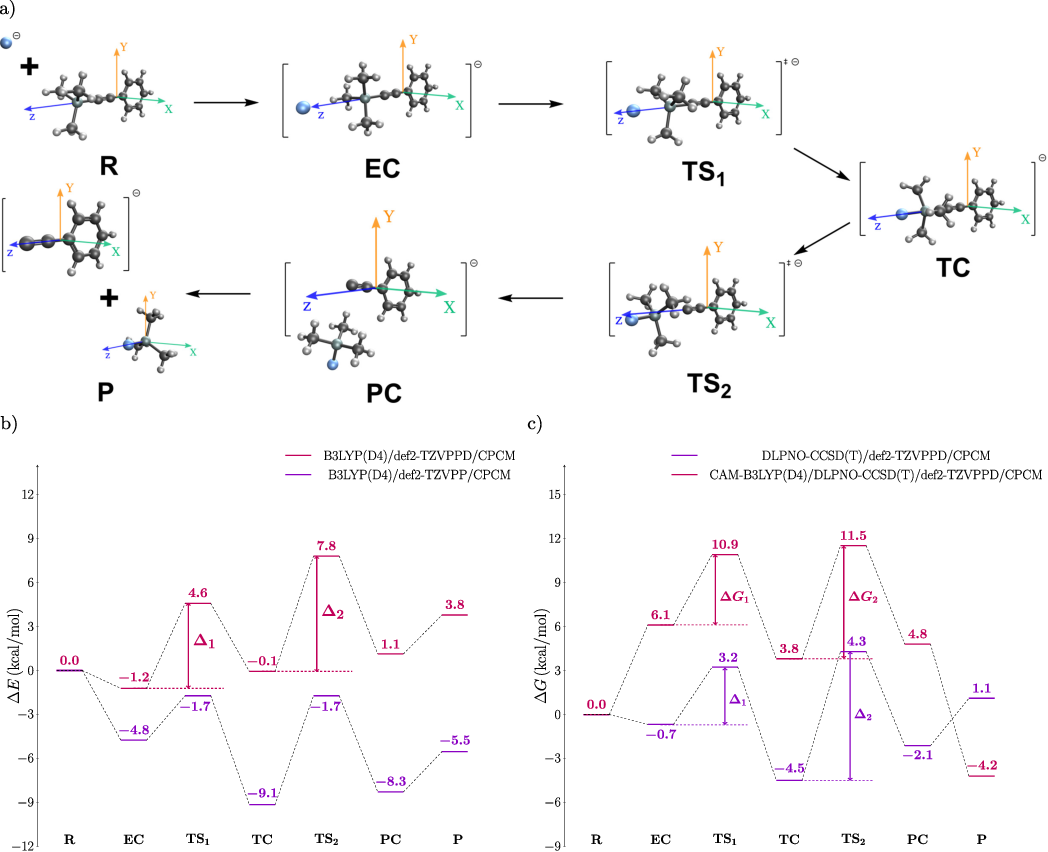}
\end{center}
\renewcommand{\baselinestretch}{1.}
\caption{Analysis of the S$_\mathrm{N}$2 reaction mechanism hypothesis. a) Stable structure of reactant (R) and products (P), encounter, product and transition complexes (EC, PC and TC) as well as transition states (TS$_1$ and TS$_2$) with species-specific body-fixed axis system diagonalizing the respective polarizability tensor. b) Electronic and activation energies, $\Delta_i,\,i=1,2$,  relative to reactants obtained on a B3LYP(D4)/CPCM level of theory for basis sets with (def2-TZVPPD) and without (def2-TZVPP) diffuse basis functions and an implicit solvation model (CPCM) for methanol. c) Electronic and activation energies relative to reactants obtained on a DLPNO-CCSD(T)/def2-TZVPPD/CPCM level of theory besides Gibbs free energies/activation free energies, $\Delta G_i,\,i=1,2$, with CC electronic components and thermodynamic corrections obtained on a CAM-B3LYP(D4)/def2-TZVPPD/CPCM level of theory. }
\label{fig.mechanism}
\end{figure*}
We first address the S$_\mathrm{N}$2 reaction mechanism with the aim of resolving discrepancies between theoretically proposed two- and single-step mechanisms on a DFT level of theory\cite{climent2020,schaefer2022}. In a second step, we refine the mechanistic hypothesis via a high-level coupled cluster treatment. All results presented in this section have been obtained via the ORCA software version 5.0.4\cite{orca2012,orcav52022}. Solvation effects of methanol were addressed implicitly via a conductor-like polarizable continuum model (CPCM) in line with previous studies\cite{climent2020,schaefer2022}. Further technical details are provided in the methodology section at the end of this work. 

We propose a two-step reaction mechanism on a dispersion-corrected B3LYP(D4)/def2-TZVPPD/CPCM level of theory with a pentacoordinate stable intermediate as shown in Figs.\ref{fig.mechanism}a and b (\textit{cf.} supporting information (SI) for molecular structures). We furthermore identify stable encounter (EC) and product complex (PC) structures, which have not been communicated so far. Since the S$_\mathrm{N}$2 reaction occurs in solution, the EC resembles the reactive reference species formed by diffusion of initially separated reactants into a stabilizing solvent cage followed by the actual chemical reaction\cite{houston2006}. In a subsequent step, we addressed the single-step mechanistic hypothesis proposed in Ref.\cite{schaefer2022}, which we were not able to reproduce on the herein given level of theory. In order to evaluate the relevance of the exchange correlation functional, we revisited the reaction mechanism on a dispersion-corrected PBE(D4)/def2-TZVPPD/CPCM level of theory but obtained a qualitatively similar two-step picture (\textit{cf.} SI).
\begin{table}[hbt!]
\caption{Activation energy, $\Delta_i$, and Gibbs free energy, $\Delta G_i$, with $i=1,2$ (\textit{cf.} Figs.\ref{fig.mechanism}b and c) relative to encounter complex (EC) obtained on B3LYP(D4)/def2-TZVPPD/CPCM and DLPNO-CCSD(T)/def2-TZVPPD/CPCM levels of theory with an implicit solvation model (CPCM) for methanol. Thermodynamic corrections were obtained on a CAM-B3LYP(D4)/def2-TZVPPD/CPCM level of theory.}
\setlength\extrarowheight{3pt}
\begin{tabular}{cl ccc}
\hline\hline
Energy $(\mathrm{kcal}/\mathrm{mol})$      && B3LYP(D4) & DLPNO-CCSD(T) & \cite{climent2020}\footnote{Energy differences relative to free reactants since encounter complex was not identified}\\
\hline
$\Delta_1$        &&   5.8  &  3.9 & 6.8 \\
$\Delta_2$        &&   7.9  &  8.8 & 8.0 \\
\vspace{-0.4cm}
\\
$\Delta G_1$      &&   6.7  &  4.8 & 14.7\\
$\Delta G_2$      &&   6.8  &  7.7 & 7.1\\
\hline\hline
\end{tabular}
\label{tab.activation_sn2}
\end{table}

We turn now to the relevance of basis sets for the quantum chemical investigation of the S$_\mathrm{N}$2 reaction. Previous studies employed relatively small Pople basis sets 6-31+G(d,p)\cite{climent2020} and 6-31G$^\star$\cite{schaefer2022} besides a larger contemporary def2-TZVP Ahlrichs basis set\cite{schaefer2024}. While all three basis sets contain additional polarization functions, only 6-31+G(d,p) includes diffuse basis functions, which are commonly exploited to accurately describe the diffuse electron density of anionic species.\cite{simons2008} Since all reactive species of the herein studied S$_\mathrm{N}$2 reaction are anionic in nature, we address the computational relevance of diffuse basis functions by comparing dispersion-corrected electronic energies (B3LYP(D4)) as shown in Fig.\ref{fig.mechanism}b obtained via basis sets def2-TZVPPD and def2-TZVPP, respectively. In absence of diffuse basis functions (def2-TZVPP), we find a stabilization of \textit{all} structures relative to the reactant energy leading qualitatively to a ``downhill'' mechanistic hypothesis. This result is contrasted by the trend obtained with diffuse basis functions (def2-TZVPPD) where only encounter (EC) and transition complexes (TC) are slightly stabilized relative to the reactants (R) while especially both transition states acquire positive relative energies. Notably, this finding is qualitatively similar to results obtained with the significantly smaller basis set 6-31+G(d,p).\cite{climent2020} We observe an identical trend with the PBE exchange correlation functional as shown in the SI and also variation of basis sets did not allow us to reproduce the single-step mechanistic hypothesis. Thus, our analysis supports the two-step mechanistic hypothesis proposed in Refs.\cite{thomas2019pta,climent2020} in agreement with concepts of S$_\mathrm{N}$2 reactions at sterically-hindered silicon centers in polar solvents\cite{bento2007,hamlin2018a,hamlin2018b}. Moreover, the inclusion of diffuse basis functions is crucial for a qualitatively correct description of anionic species in the S$_\mathrm{N}$2 reaction with direct consequences for the quality of potential energy surfaces employed in subsequent vibrational or vibro-polaritonic studies.

We now discuss the energetics of the proposed two-step mechanism in closer detail. To this end, we additionally exploit the ``gold standard of single-reference quantum chemistry'' in terms of coupled cluster theory with singles, doubles and perturbative triples (CCSD(T)) in the efficient domain-based local pair natural orbital (DLPNO) implementation.\cite{riplinger2013a,riplinger2013b} In Fig.\ref{fig.mechanism}c, we show electronic energies on a DLPNO-CCSD(T)/def2-TZVPPD/CPCM level of theory and find all reactive species to be stabilized relative to previous DFT results shown in Fig.\ref{fig.mechanism}b. In Tab.\ref{tab.activation_sn2}, we compare activation energies obtained here and in Ref.\cite{climent2020}, where we like to emphasize that the first barrier is here characterized via the newly identified EC in contrast to Ref.\cite{climent2020}, which takes the free reactants (R) as energetic reference. In general, we recover the trend of the initial barrier, $\Delta_1$, being smaller than the second barrier, $\Delta_2$, via both DFT and CC methodologies, such that the second step determines the reaction kinetics. However, $\Delta_1$ is significantly reduced on a CC level of theory, while $\Delta_2$ varies only slightly between DFT and CC results. This finding is rationalized via a significant stabilization of the first transition state structure accompanied by a slight destabilization of the encounter complex on a CC level of theory. 
Gibbs free energies were subsequently obtained via thermodynamic corrections on a CAM-B3LYP(D4)/def2-TZVPPD/CPCM level of theory. We observe here a significant thermodynamic stabilization of reactant and product structures, which can be attributed to missing translational entropy contributions when turning from two separated particles to a single entity. We furthermore note that the computational entropy correction relies on the free particle model and is therefore overestimated in the herein discussed condensed phase setting. Turning to Gibbs free energies, we find the first barrier as related to the activation free energy, $\Delta G_1$, from the EC to the TS$_1$, to be smaller than the second barrier, $\Delta G_2$. Accordingly, the energetic ordering of Gibbs free energy barriers is similar to their purely electronic equivalent with the second step being central for the reaction kinetics.

\subsection{Electronic Dipole Fluctuation Corrections}
\label{sec.dip_fluc_correct}
We are now in the position to address cavity-induced electronic corrections of the molecular PES underlying the two-step mechanistic hypothesis discussed before. In the cavity Born-Oppenheimer framework\cite{flick2017,flick2017cbo,fischer2023}, electronic energies are subject to dipole fluctuation corrections under VSC
\begin{align}
\Delta^{(ec)}_\lambda
&=
\dfrac{g^2_0}{2}
\braket{
\Psi_0
\vert
\left(
\Delta
\hat{d}^{(e)}_\lambda
\right)^2
\vert
\Psi_0}
\propto
\dfrac{g^2_0}{2}
\alpha_{\lambda\lambda}
\quad,
\label{eq.unsoeld_polarize}
\end{align}
with $\Delta\hat{d}^{(e)}_\lambda=\hat{d}^{(e)}_\lambda-\braket{\Psi_0\vert\hat{d}^{(e)}_\lambda\vert\Psi_0}$, where $\hat{d}^{(e)}_\lambda$ is the polarization-projected electronic dipole operator for a cavity mode with polarization, $\lambda$, and $\ket{\Psi_0}$ is the adiabatic ground state.\cite{fischer2024ele} The dipole fluctuation correction, $\Delta^{(ec)}_\lambda$, can be conceptually connected with a diagonal element of the polarizability tensor, $\alpha_{\lambda\lambda}$, by exploiting the Uns\"old approximation\cite{unsold1927,sylvain1987,hait2023}, which illustrates a recent proposal on the potential relevance of electronic polarizability effects under VSC.\cite{schaefer2024}

In the following, we systematically address cavity-induced dipole fluctuation corrections of DLPNO-CCSD(T) results discussed in Sec.\ref{sec.mechanist_hypo} from a non-perturbative \textit{ab initio} perspective. Specifically, we exploit CBO Hartree-Fock and coupled cluster methods in their cavity reaction potential (CRP) formulations, CRP-HF and lCRP-CCSD,\cite{fischer2024ele,fischer2025} as recently implemented via capabilities of the Python-based Simulations of Chemistry Framework (PySCF) package\cite{sun2018,sun2020}. This approach allows us to obtain dipole fluctuation corrections on both mean-field and correlated levels of theory (\textit{cf.} methodology section for details). Moreover, we account for cavity-induced molecular reorientation by choosing the molecular body-fixed axis system of individual reactive species such that the respective polarizability tensor is diagonal (\textit{cf.} Fig.\ref{fig.mechanism}a).\cite{schnappinger2024struc} In this context, Eq.\eqref{eq.unsoeld_polarize} indicates an increase of the dipole fluctuation correction for $x$- to $z$-polarized cavity modes simply through the ordering of polarizability tensor eigenvalues. This trend is illustrated by molecular structures shown in Fig.\ref{fig.mechanism}a, where the $z$-axis lies in in the molecular plane parallel to the the easily polarizeable Si-C-bond and phenyl-system.  
\begin{figure}[hbt!]
\begin{center}
\includegraphics[scale=1.0]{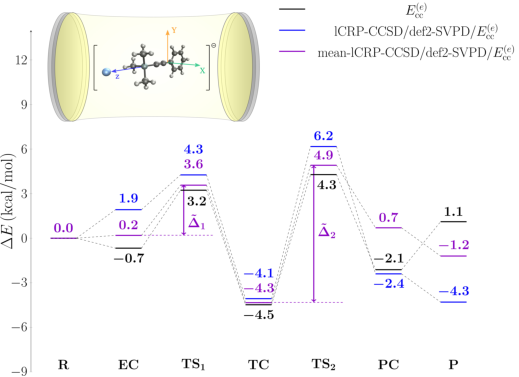}
\end{center}
\renewcommand{\baselinestretch}{1.}
\caption{Electronic energies on DLPNO-CCSD(T)/def2-TVZPPD/CPCM level of theory, $E^{(e)}_\mathrm{cc}$, relative to reactants corrected by cavity-induced dipole fluctuation contributions obtained on lCRP-CCSD/def2-SVPD level of theory for a $z$-polarized cavity mode and mean-value equally accounting for $x$-, $y$- and $z$-polarization.}
\label{fig.dipole_fluctuation_effects}
\end{figure}

We consider dipole fluctuation corrections for all reactive species discussed in Fig.\ref{fig.mechanism} for light-matter interaction strength, $g_0=0.015\,\sqrt{E_h}/ea_0$. In line with previous observations\cite{fischer2024ele,fischer2025}, we find corrections on a mean-field level (CRP-HF) to be significantly smaller than their correlated counterparts (lCRP-CCSD) (\textit{cf.} SI). We thus focus on correlated dipole fluctuation corrections, which are shown for different cavity-reoriented reactive species of the S$_\mathrm{N}$2 reaction in Tab.\ref{tab.dipole_fluctuation_details} relative to the corrected reactant energies for different choices of cavity mode polarization. 
\begin{table}[hbt!]
\caption{Cavity-induced electronic dipole fluctuation corrections relative to reactants (R) in kcal/mol obtained on lCRP-CCSD/def2-SVPD level of theory for different cavity polarizations, $\lambda$, with light-matter interaction strength, $g_0=0.015\,\sqrt{E_h}/ea_0$.}
\setlength\extrarowheight{3pt}
\begin{tabular}{cccccccccccc}
\hline\hline
$\lambda$ & EC && TS$_1$ && TC && TS$_2$ &&  PC && P \\
\hline
$x$                 &  0.01  && 0.04  && 0.07    && 0.08 &&  0.34  &&  0.24   \\
$y$                 &  0.01  && 0.05  && 0.08    && 0.07 &&  4.01  &&  0.49   \\
$z$                 &  2.59  && 1.03  && 0.41    && 1.89 && -2.40  && -4.33   \\
$\frac{x+y+z}{3}$   &  0.86  && 0.34  && 0.14    && 0.63 &&  0.65  && -1.2    \vspace{0.1cm}\\
\hline\hline
\end{tabular}
\label{tab.dipole_fluctuation_details}
\end{table}

We observe dipole fluctuation corrections to be particularly pronounced for the $z$-polarization scenario in line with expectations deduced from Eq.\eqref{eq.unsoeld_polarize}. In order to mimic condensed phase environments potentially preventing ideal molecular alignment, we account for a forth polarization scenario with equal contributions from all molecular axis. In Fig.\ref{fig.dipole_fluctuation_effects}, we depict the dipole-fluctuation corrected two-step mechanism (lCRP-CCSD/def2-SVPD) with electronic reference on a DLPNO-CCSD(T)/def2-TVZPPD/CPCM level of theory for the average and $z$-polarization scenario. 
We observe the dipole fluctuation correction to destabilize all reactive species from the encounter complex to the second transition state, which results in a slightly smaller first barrier, $\tilde{\Delta}_1=3.5\,\mathrm{kcal}/\mathrm{mol}$, and a slightly larger second barrier, $\tilde{\Delta}_2=9.2\,\mathrm{kcal}/\mathrm{mol}$, under VSC (\textit{cf.} Tab.\ref{tab.activation_sn2}) affecting the rate determining reaction step. Product complex (PC) and product (P) species are found to behave rather differently, which can be explained via their significantly altered molecular polarizability properties resulting from the cleavage of the Si-C bond. We thus conclude that the S$_\mathrm{N}$2 reaction model of PTA under VSC theoretically exhibits non-negligible dipole fluctuation corrections on a correlated level of theory, which have not been discussed so far but are qualitatively in agreement with previous reactive model studies\cite{fischer2025}. 

We close this section by noting that our study provides only a first step towards a microscopic description of vibro-polaritonic chemistry since relevant cavity-induced collective effects and solvent effects are currently beyond the capabilities of the methodology exploited in this study. Moreover, the connection between the experimentally central cavity resonance effect and theoretically discussed cavity-induced electronic corrections constitutes another relevant open question.

\subsection{Si-C-Stretching Motion from a Dipole Perspective}
\begin{figure*}[hbt!]
\begin{center}
\includegraphics[scale=1.0]{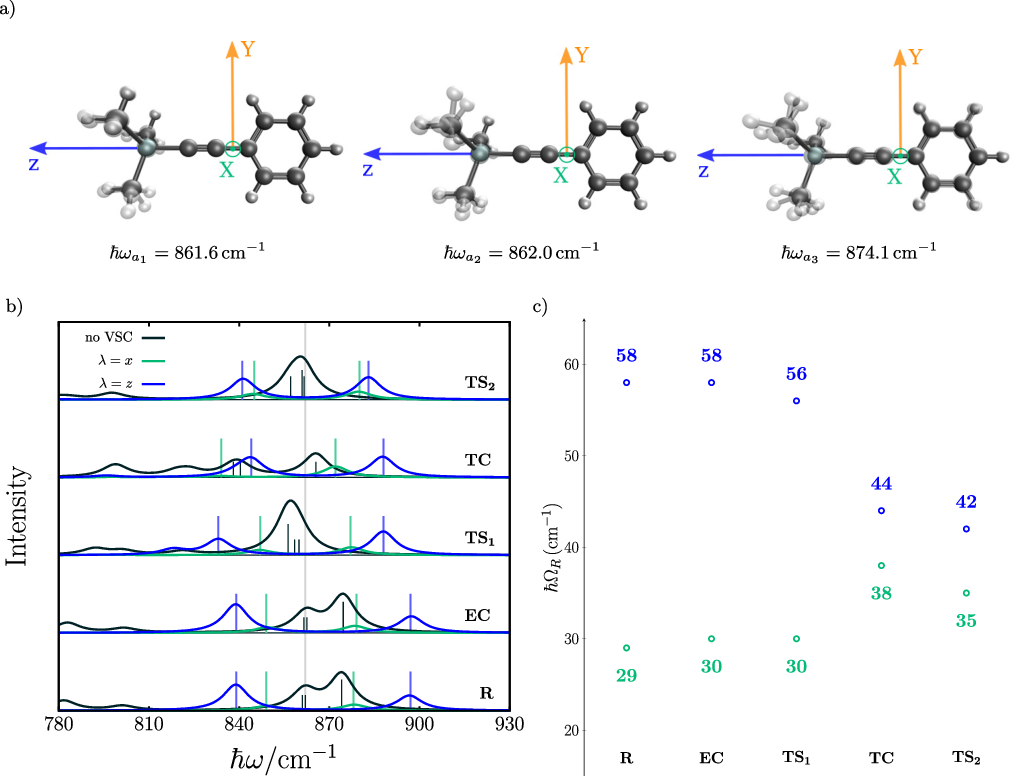}
\end{center}
\renewcommand{\baselinestretch}{1.}
\caption{a) Normal modes contributing to the double-peak feature of PTA with (from left to right) two CH$_3$-rocking modes and a mode of mixed CH$_3$-rocking/Si-C-stretching character. b) Theoretical linear IR spectra for different reactive species containing the relevant Si-C-bond and related effective cavity transmission spectra for VSC involving a single cavity mode at frequency $862\,\mathrm{cm}^{-1}$ (vertical line) for polarizations $\lambda=x$ (green) and $\lambda=z$ (blue) at light-matter interaction strength, $g_0=0.015\,\sqrt{E_h}/ea_0$. Effective transmission spectra were obtained via the CBO-PT(2) linear response approach and vibrational polariton peaks are indicated by vertical bars. c) Rabi splittings for polarizations $\lambda=x$ (green) and $\lambda=z$ (blue) as function of reactive species obtained via the CBO-PT(2) linear response approach.}
\label{fig.vsc_effects}
\end{figure*}
We shall finally reconsider the vibrational double-peak feature around $860\,\mathrm{cm}^{-1}$ in the linear IR spectrum of PTA with a focus on Si-C-stretching motion.\cite{thomas2016,climent2020} In contrast to previous theoretical work\cite{climent2020}, we will analyse the relevant modes from the perspective of their dipole character, which is directly related to both linear IR response and vibrational polariton formation. We recall that all molecular structures discussed here account for cavity-induced reorientation as before, such that the relevant Si-C-bond is parallel to the $z$-axis of the respective coordinate system (\textit{cf.} Fig.\ref{fig.mechanism}a).

We start our discussion with the results of a canonical normal-mode analysis for PTA on a CAM-B3LYP(D4)/def2-TZVPPD/CPCM level of theory, where we recover previously discussed normal modes in the energetically relevant region\cite{climent2020}: Two nearly degenerate modes at $862\,\mathrm{cm}^{-1}$ with dominant CH$_3$-rocking character ($a_1,a_2$) and a single mode at $874\,\mathrm{cm}^{-1}$ $(a_3)$, which additionally exhibits a significant Si-C-stretching contribution (\textit{cf.} Fig.\ref{fig.vsc_effects}a). We characterize this Si-C-stretching contribution via the relative displacement, $\Delta r_z$, of the Si-C-bond along the $z$-axis in line with Ref.\cite{climent2020} 
\begin{align}
\Delta r_z
&=
\vert
\Delta\underline{r}_\mathrm{Si}
-
\Delta\underline{r}_\mathrm{C}
\vert
\cdot
\underline{e}_z
\quad,
\end{align}
with $\Delta\underline{r}_i=\underline{r}^i_\mathrm{max}-\underline{r}^i_\mathrm{min}$ for $i=\mathrm{Si},\mathrm{C}$, which is effectively zero for CH$_3$-rocking modes $a_1$ and $a_2$ at the herein discussed level of accuracy as shown in Tab.\ref{tab.vib_details_pta}. 
\begin{table}[hbt!]
\caption{Vibrational properties of PTA in principal axis frame of its polarizability tensor obtained on CAM-B3LYP(D4)/def2-TZVPPD/CPCM level of theory. Normal-mode frequencies, $\tilde{\nu}$, relative displacement of Si-C bond along $z$-axis, $\Delta r_z$, intensity, $I\propto\sum_\kappa T^2_\kappa$, frequency-weighted dipole derivatives, $T_\kappa=\frac{1}{\sqrt{2\omega_i}}\frac{\partial d_\kappa}{\partial Q_i}$, along Cartesian coordinates, $\kappa=x,y,z$, with $x$-axis perpendicular to the molecular plane and $z$-axis parallel to the Si-C bond.}
\setlength\extrarowheight{3pt}
\begin{tabular}{ccccccccccc}
\hline\hline
Mode & $\tilde{\nu}/\mathrm{cm}^{-1}$ & $\Delta r_z/\mathrm{a.u.}$ && $I/\mathrm{a.u.}$ && $T_x/\mathrm{a.u.}$ && $T_y/\mathrm{a.u.}$ &&  $T_z/\mathrm{a.u.}$ \\
\hline
$a_1$  &  861.6  & ---   && 0.01    && -0.09 && -0.06  &&  ---  \\
$a_2$  &  862.0  & ---   && 0.01    &&  0.05 && -0.09  &&  ---  \\
$a_3$  &  874.1  & 0.13  && 0.04    &&  ---  &&  ---   && -0.19 \\
\hline\hline
\end{tabular}
\label{tab.vib_details_pta}
\end{table}

We shall turn our attention now to the spectroscopic relevance of the $a_3$-mode's Si-C-stretching contribution motivated by the observation that the related intensity, $I$, is significantly higher compared to the two CH$_3$-rocking modes $a_1$ and $a_2$ (\textit{cf.} Tab.\ref{tab.vib_details_pta}). Since the intensity is directly related to Cartesian components of transition dipole derivatives, $T_\kappa$, we can rationalize this finding via a dominant contribution along the $z$-axis reflecting the Si-C-stretching motion of the $a_3$ mode. Moreover, similar contributions are negligible for modes $a_1$ and $a_2$, with CH$_3$-rocking motion being dominantly active along $x$- and $y$-axis. We may thus conclude, that only the $a_3$-mode contains a significant Si-C-stretching contribution despite the presence of additional CH$_3$-rocking components as shown in Fig.\ref{fig.vsc_effects}a.
\begin{table}[hbt!]
\caption{Energies and mode-resolved probabilities of lower and upper vibrational polaritons, $\ket{\mathrm{LP}(\lambda)}$ and $\ket{\mathrm{UP}(\lambda)}$, for a four-dimensional model of the spectroscopic double-peak feature of PTA containing two CH$_3$-rocking modes $(a_1,a_2)$, the mixed Si-C stretching mode ($a_3$) and a single cavity mode $(a_\mathrm{cav})$ with polarization, $\lambda=x,z$, at light-matter interaction strength, $g_0=0.015\,\sqrt{E_h}/ea_0$.}
\setlength\extrarowheight{3pt}
\begin{tabular}{cccccccccccc}
\hline\hline
State      && $\tilde{\nu}/\mathrm{cm}^{-1}$ && $\vert a_1\vert^2$ && $\vert a_2\vert^2$ &&  $\vert a_3\vert^2$ && $\vert a_\mathrm{cav}\vert^2$ \\
\hline
$\ket{\mathrm{LP}(x)}$       && 847 &&    0.35 &&  0.14  && --- &&  0.51 \\
$\ket{\mathrm{UP}(x)}$       && 877 &&    0.35 &&  0.14  && --- &&  0.51 \\
\vspace{-0.4cm}
\\
$\ket{\mathrm{LP}(z)}$       && 839 &&     ---   &&  ---  &&  0.36 &&  0.64 \\
$\ket{\mathrm{UP}(z)}$       && 897 &&     ---   &&  ---  &&  0.64 &&  0.36 \\
\hline\hline
\end{tabular}
\label{tab.vib_polariton_details}
\end{table}

As the dipole character of a vibrational (normal) mode directly manifests in its relevance for vibrational polariton formation, this observation suggests that the $a_3$-mode is central for VSC effects in the spectral region of the double-peak feature of PTA. In order to illustrate this hypothesis, we consider a reduced four-mode model containing three normal modes $a_1,a_2$ and $a_3$, besides a single cavity mode with frequency, $\omega_c=862\,\mathrm{cm}^{-1}$. We obtain lower and upper vibrational polaritons (LP and UP) via the CBO-PT(2) approach\cite{fischer2024vib} (\textit{cf.} methodology section) for two different cavity polarizations, $\lambda=x,z$ (perpendicular and parallel to the Si-C bond), at light-matter interaction strength, $g_0=0.015\,\sqrt{E_h}/ea_0$. Corresponding energies and zero-order state probabilities are given in Tab.\ref{tab.vib_polariton_details} and effective transmission spectra are shown in Fig.\ref{fig.vsc_effects}b (bottom). 
We find that $x$-polarization (perpendicular to the Si-C bond) leads to LP- and UP-states with pronounced $a_1$- and $a_2$-contributions corresponding to a superposition of CH$_3$-rocking modes, whereas the $a_3$-component is negligible in this scenario. We note that a $y$-polarized cavity mode leads to a similar scenario as $x$-polarization with interchanged contributions from $a_1$- and $a_2$-modes and is therefore not shown here. In contrast, we find for a $z$-polarized cavity mode (parallel to the Si-C bond) a dominant contribution of the mixed Si-C-stretching mode, $a_3$, in vibrational polariton formation, which accurately reflects the trend indicated by transition dipole derivative components shown in Tab.\ref{tab.vib_details_pta}. 

We turn now to the Rabi splittings for the two different polarization scenarios and find VSC in the $z$-polarization scenario with $\hbar\Omega_z=58\,\mathrm{cm}^{-1}$ to be significantly stronger compared to the $x$-polarization scenario with $\hbar\Omega_x=29\,\mathrm{cm}^{-1}$ (\textit{cf.} Fig.\ref{fig.vsc_effects}b, bottom). Since the Rabi splitting is directly proportional to the transition dipole derivative of the strongly coupled mode in a normal-mode scenario, this result is again rationalized by the dominant dipole component of the mixed Si-C stretching mode along the $z$-axis. 
Accordingly, our theoretical analysis suggests that the Si-C-stretching contribution of the $a_3$-mode is indeed central for both linear IR response and vibrational polariton formation in agreement with experiment\cite{thomas2016} despite (partial) CH$_3$-rocking contributions to all three normal modes forming the double-peak feature. We furthermore note that the presented polarization scenario is of course an ideal limit and realistically vibrational polaritons will contain contributions from all three modes, which however does not alter the relevance of the Si-C stretching component. 

Since the relevant Si-C-bond of PTA is cleaved throughout the S$_\mathrm{N}$2 reaction, we will eventually discuss the evolution of the Si-C-stretching contribution up to the second transition state (TS$_2$). In Figs.\ref{fig.vsc_effects}b and \ref{fig.vsc_effects}c, we show molecular IR spectra and their cavity response equivalent besides Rabi splittings for different reactive species obtained at the previously fixed cavity frequency, $\omega_c=862\,\mathrm{cm}^{-1}$. Throughout the reaction, we observe a red-shift of the double-peak feature accompanied by a decrease of the Rabi splitting, $\hbar\Omega_z$, for the $z$-polarization scenario. In order to explain this trend, we extended the normal-mode analysis of PTA (\textit{cf.} Tab.\ref{tab.vib_details_pta}) to the other reactive species considered here (\textit{cf.} SI) and observe a corresponding decrease of the dipole derivative component along the $z$-axis. We may now recall that the Si-C-bond is aligned parallel to the $z$-axis for all relevant compounds as shown in Fig.\ref{fig.mechanism}a, which rationalizes the decreasing Rabi splitting, $\hbar\Omega_z$, with the related changes in dipole character during the Si-C-bond cleavage process. Specifically, the Rabi splitting decreases in line with a decrease of the dipole's $z$-component along the Si-C-bonding axis (\textit{cf.} SI). Structural changes throughout the reaction lead additionally to a slight increase of the Rabi splitting for the $x$-polarization scenario.  

In summary, we find a strong correlation between the Si-C-stretching motion of the $a_3$-mode and its dipole component along the Si-C-bond, which manifests in its dominant IR response and contribution to vibrational polariton formation. Thus, by correlating structural and optical properties of normal modes, we conclude that the vibrational double-peak feature of PTA indeed exhibits a prominent Si-C-stretching component relevant for VSC in agreement with experiment despite significant contributions of CH$_3$-rocking motion.

\section{Conclusions}
\label{sec.conclusion}
We theoretically investigated the S$_\mathrm{N}$2 reaction of PTA with fluoride under VSC from a quantum chemical \textit{ab initio} perspective with a focus on three complementary aspects: A refined mechanistic hypothesis, cavity-induced corrections of electronic energies under VSC and the optical relevance of a Si-C-stretching motion in the vibrational double-peak feature of PTA. 

In the first part, we reported on computational evidence for a two-step mechanism in agreement with previous suggestions\cite{thomas2019pta,climent2020} based on density functional theory and high-level coupled cluster theory. We furthermore identified stable encounter and product complex structures, which were not communicated so far, with the encounter complex providing the reactive reference species of the S$_\mathrm{N}$2 reaction in solution. We additionally showed that diffuse basis functions are crucial for a qualitatively correct quantum chemical description of anionic species in the S$_\mathrm{N}$2 reaction. We were not able to reproduce a previously reported alternative single-step mechanism\cite{schaefer2022} but obtained qualitatively the same two-step picture for different exchange correlation functionals and basis sets.

We subsequently addressed cavity-induced dipole fluctuation corrections for the two-step mechanism by exploiting recent implementations of non-perturbative CBO Hartree-Fock and coupled cluster theories.\cite{fischer2024ele,fischer2025} We observe significant correlated corrections in the single-molecule limit for a coupled cluster treatment, which were most pronounced for a cavity-mode polarization parallel to the largest component of the molecular polarizabilty tensor. A connection between electronic dipole fluctuation corrections and electronic polarizability components was qualitatively established via the Uns\"old approximation, which extends a recent suggestion on cavity-induced molecular reorientation based on properties of the polarizability tensor\cite{schnappinger2024struc} by an energetic argument. Our \textit{ab initio} results support furthermore a previously stated hypothesis\cite{schaefer2024} on the relevance of electronic polarizability effects under VSC. 

In the third and final part of our work, we analysed the relevance of CH$_3$-rocking and Si-C-stretching motion of PTA under VSC from the perspective of normal modes' dipole character. We find that the Si-C-stretching motion is characterized by a strong dipole component along the Si-C-bond, which renders it dominant for both linear IR response and vibrational polariton formation. This connection was furthermore indicative for the evolution of Rabi splittings throughout the proposed two-step mechanism. Our analysis suggests that it is indeed the Si-C-stretching contribution that is dominant in the VSC process in line with experimental arguments despite a clear assignment as ``pure'' Si-C-stretching mode might be difficult due to the complexity of the molecule manifesting in additional CH$_3$-rocking contributions. 

Our work highlights the capabilities of recent developments in \textit{ab initio} vibro-polaritonic chemistry.\cite{schnappinger2023,fischer2024vib,schnappinger2024struc,fischer2024ele,fischer2025} but constitutes eventually only a next step towards the complexity of vibro-polaritonic chemistry under experimental conditions. Specifically solvent effects under VSC as well as the crucial but theoretically challenging collective strong coupling regime remain pressing open issues relevant for reducing the conceptual gap between theory and experiment further.

\section*{Methodology}
\label{sec.method}
\textit{Quantum Chemistry}. All quantum chemical calculations were performed via the ORCA software package\cite{orca2012} in version 5.0.4\cite{orcav52022}. In a first step, we reoptimized the pentacoordinate transition complex structure reported in both Refs.\cite{climent2020,schaefer2022} via the hybrid functional B3LYP and the D4-dispersion correction\cite{caldeweyher2017,caldeweyher2019} with a large contemporary Ahlrich-type basis set def2-TZVPPD containing additional polarization and diffuse basis functions\cite{weigend2005,pantazis2012}. Implicit solvent effects were taken into account via the conductor-like polarizable continuum model (CPCM) for methanol as implemented in ORCA 5.0.4.\cite{garciarates2020} We subsequently identified transition states, encounter and product complex species via an intrinsic reaction coordinate (IRC) approach.\cite{morokuma1977} Stationary points on the molecular potential energy surface were characterized via a normal mode analysis of all reactive species. 
Electronic energies were obtained for all optimized structures on B3LYP(D4)/def2-TZVPPD/CPCM and DLPNO-CCSD(T)/def2-TZVPPD/CPCM levels of theory with a CPCM for methanol. We investigated the role of diffuse basis functions by comparing electronic energies obtained from def2-TVZPP and def2-TZVPPD basis sets.
Molecular normal modes, dipole moment derivatives and polarizability tensors were obtained via the hybrid functional CAM-B3LYP\cite{yanai2004} on a CAM-B3LYP(D4)/def2-TZVPPD/CPCM level of theory for cavity-reoriented molecular structures as discussed below.

\textit{Polaritonic Chemistry}. Cavity-induced molecular reorientation is approximately accounted for by transforming molecular structures to the principal axis frame of the polarizability tensor\cite{schnappinger2024struc}, which provides a molecule-specific unique way of choosing cavity-mode polarization vectors. Diagonal polarizability tensors and their eigenvectors, which provide a transformation to the polarizability tensor's principal axis frame, have been numerically obtained on CAM-B3LYP(D4)/def2-TZVPPD/CPCM level of theory as discussed above. Further cavity-induced relaxation effects of internal molecular coordinates have been recently identified to be small in the single-molecule limit and are therefore neglected here.\cite{schnappinger2024struc,liebenthal2024,lexander2024}

We addressed cavity-induced electronic corrections from the perspective of recently developed \textit{ab initio} CBO wave function approaches on both mean-field and correlated levels of theory, specifically, CBO restricted Hartree-Fock (CBO-RHF) and CBO coupled cluster theories (CBO-CC) in the cavity reaction potential (CRP) formulation.\cite{fischer2024ele,fischer2025}. We exploit here a Python-based implementation of both CRP-RHF and CRP-CCSD methods realized via the Python-based Simulations of Chemistry Framework (PySCF) package.\cite{sun2018,sun2020} 
The CRP approach provides a numerical route towards the ground state CBO electronic energy 
\begin{align}
E^{(ec)}_0(\underline{R})
&=
\mathcal{V}^{(ec)}_\mathrm{rhf}(\underline{R})
+
\Delta\mathcal{V}^{(ec)}_\mathrm{corr}(\underline{R})
\quad,
\label{eq.ground_state_cpes_crp}
\end{align}
minimized in cavity coordinate space with (mean-field) CRP-RHF energy
\begin{align}
\mathcal{V}^{(ec)}_\mathrm{rhf}
=
E^{(ec)}_\mathrm{rhf}
+
g^2_0
\left(
\sum_i
O^{ii}_\lambda
-
\sum_{ij}
d^{ij}_\lambda
d^{ji}_\lambda
\right)
\quad.
\label{eq.mean_field_crp}
\end{align}
The first term reflects the RHF energy, $E^{(ec)}_\mathrm{rhf}$, evaluated with molecular orbitals (MOs) relaxed by the electronic dipole self-energy, and the second contribution resembles the mean-field electronic dipole fluctuation correction determined by electronic dipole and quadrupole matrix elements, $d^{ij}_\lambda$ and $O^{ii}_\lambda$, with cavity mode polarization, $\lambda$.\cite{fischer2025} Indices $i,j$ run over the occupied MO subspace. The second term in Eq.\eqref{eq.ground_state_cpes_crp} constitutes the cavity-modified electron correlation correction, which reads for the linearized CRP-CC approach accounting for singles and doubles excitations (lCRP-CCSD)\cite{fischer2025}  
\begin{multline}
\Delta\mathcal{V}^{(ec)}_\mathrm{corr}
=
\sum_{aibj}
\left(
t^a_i
t^b_j
+
t^{ab}_{ij}
\right)
\left(
2g_{iajb}
-
g_{ibja}
\right)
\\
+
g^2_0
\sum_{aibj}
\left(
2
t^{ab}_{ij}
d^{ai}_\lambda
d^{bj}_\lambda
-
\left(
t^a_i
t^b_j
+
t^{ab}_{ij}
\right)
d^{bi}_\lambda
d^{aj}_\lambda
\right)
\quad.
\label{eq.crp_ccsd_correlation_energy}
\end{multline}
The first line reflects the canonical CCSD correlation energy for closed-shell systems with singles and doubles amplitudes, $t^a_i, t^b_j$ and $t^{ab}_{ij}$, relaxed by the electronic dipole self-energy besides electron repulsion integrals, $g_{iajb}$. The second line resembles the correlation contribution to the electronic dipole fluctuation correction with polarization-projected transition dipole matrix elements, $d^{ai}_\lambda$ and $d^{bj}_\lambda$, in the MO basis. Indices $a,b$ run here over the virtual MO subspace. Moreover, Eq.\eqref{eq.crp_ccsd_correlation_energy} specifically corresponds to the $\Lambda_0$-\textit{linearized} approximation of the CRP-CCSD correlation energy, which accounts for additional correlation corrections to the stationary cavity coordinate, $x^0_\lambda$.\cite{fischer2025} 

Vibrational polaritons and their linear IR spectra are addressed via the linear response formulation of second-order CBO perturbation theory (CBO-PT)\cite{fischer2024vib}, which perturbatively accounts for non-resonant interactions between electrons and cavity modes in the VSC regime. This interaction manifests in a second-order cavity contribution to the linear IR spectrum interpretable as effective transmission spectrum. The corresponding vibro-polaritonic IR spectrum 
\begin{align}
\sigma_\mathrm{cav}(\hbar\omega)
&=
\sum^{N_p}_m
I^m_\mathrm{cav}
\delta
\left(
\hbar\omega
-
\hbar\Omega_m
\right)
\quad,
\end{align}
is determined by $N_p$ vibrational polariton energies, $\hbar\Omega_m$, obtained from the respective second-order Hessian and a cavity intensity component, $I^m_\mathrm{cav}$, which have both explicitly been given in Ref.\cite{fischer2024vib} (\textit{cf.} Eqs.(23)-(27) therein). The linear response CBO-PT(2) approach is realized via normal modes, dipole derivatives and polarizability tensor elements as obtained from quantum chemistry calculations discussed above.

\section*{Supplementary Information}
Contains additional details on the refined mechanistic hypothesis, electronic dipole fluctuation corrections, normal modes as well as optimized and cavity-reoriented molecular structures in Cartesian coordinates.

\section*{Conflict of Interest}
The author has no conflicts to disclose.

\section*{Data Availability Statement}
The data supporting this article have been included as part of the Supplementary Information.

\section*{Acknowledgements}
E.W. Fischer acknowledges funding by the Deutsche Forschungsgemeinschaft (DFG, German Research Foundation) through DFG project 536826332 and helpful discussions with Philipp Woite, Leon Gerndt and Muhammed B\"uy\"uktemiz.


\onecolumngrid
\subsection*{Supplementary Information -- Nucleophilic substitution at silicon under vibrational strong coupling: Refined insights from a high-level \textit{ab initio} perspective}

\subsection*{Details on Refined Mechanistic Hypothesis}
\begin{figure*}[hbt!]
\begin{center}
\includegraphics[scale=0.5]{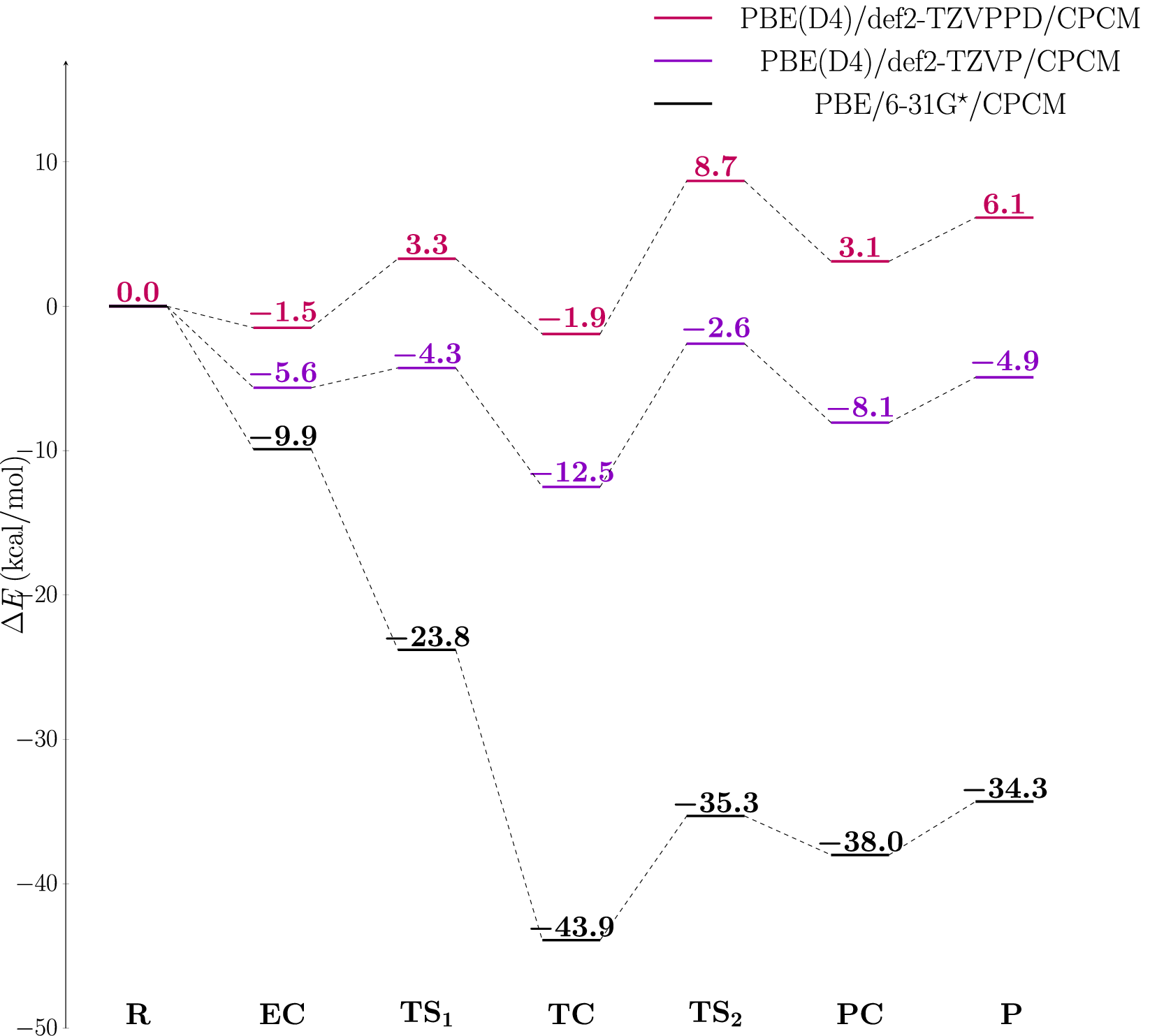}
\end{center}
\renewcommand{\baselinestretch}{1.}
\caption{Electronic energies of the S$_\mathrm{N}$2 reaction mechanism hypothesis obtained on PBE(D4)/def2-TZVPPD/CPCM, PBE(D4)/def2-TZVP/CPCM and PBE/6-31G$^\star$/CPCM levels of theory with an implicit solvation model (CPCM) for methanol. Underlying molecular structures were optimized on B3LYP(D4)/def2-TZVPPD/CPCM level of theory.}
\label{fig.pbe_mechanism}
\end{figure*}

\newpage

\subsection*{Details on Dipole Fluctuation Corrections}
\begin{table}[hbt!]
\caption{Absolute cavity-induced electronic dipole fluctuation corrections in kcal/mol obtained on CRP-HF/def2-SVPD level of theory for different cavity polarizations, $\lambda$, with light-matter interaction strength, $g_0=0.015\,\sqrt{E_h}/ea_0$.}
\setlength\extrarowheight{3pt}
\begin{tabular}{cccccccccccccc}
\hline\hline
$\lambda$   & R && EC && TS$_1$ && TC && TS$_2$ &&  PC && P \\
\hline
$x$                 &  3.73  &&  3.74  &&  3.75   &&  3.75 &&  3.79  && 3.78  && 3.77  \\
$y$                 &  4.17  &&  4.18  &&  4.20   &&  4.19 &&  4.28  && 4.20  && 4.21 \\
$z$                 &  4.59  &&  4.58  &&  4.58   &&  4.59 &&  4.61  && 4.60  && 4.60  \\
\hline\hline
\end{tabular}
\label{tab.abs_dipole_fluctuation_crprhf}
\end{table}
\begin{table}[hbt!]
\caption{Cavity-induced electronic dipole fluctuation corrections relative to reactants (R) in kcal/mol obtained on CRP-HF/def2-SVPD level of theory for different cavity polarizations, $\lambda$, with light-matter interaction strength, $g_0=0.015\,\sqrt{E_h}/ea_0$.}
\setlength\extrarowheight{3pt}
\begin{tabular}{cccccccccccc}
\hline\hline
$\lambda$  & EC && TS$_1$ && TC && TS$_2$ &&  PC && P \\
\hline
$x$                 &  0.01   &&  0.02   &&  0.02  &&  0.06  && 0.06  && 0.05  \\
$y$                 &  0.01   &&  0.04   &&  0.02  &&  0.11  && 0.03  && 0.04 \\
$z$                 &  -0.02  && -0.01   && -0.002 &&  0.01  && 0.01  && 0.01 \\
\hline\hline
\end{tabular}
\label{tab.rel_dipole_fluctuation_crprhf}
\end{table}
\begin{table}[hbt!]
\caption{Absolute cavity-induced electronic dipole fluctuation corrections in kcal/mol obtained on lCRP-CCSD/def2-SVPD level of theory for different cavity polarizations, $\lambda$, with light-matter interaction strength, $g_0=0.015\,\sqrt{E_h}/ea_0$.}
\setlength\extrarowheight{3pt}
\begin{tabular}{cccccccccccccc}
\hline\hline
$\lambda$  & R && EC && TS$_1$ && TC && TS$_2$ &&  PC && P \\
\hline
$x$                 &  3.74  &&  3.75  &&  3.79   &&  3.81 &&  3.82  && 4.08  && 3.98  \\
$y$                 &  4.50  &&  4.51  &&  4.55   &&  4.58 &&  4.57  && 8.59  && 4.99 \\
$z$                 & 10.86  && 13.45  && 11.89   && 11.27 && 12.74  && 8.46  && 6.53  \\
\hline\hline
\end{tabular}
\label{tab.abs_dipole_fluctuation_crpccsd}
\end{table}

\newpage 

\subsection*{Details on Normal Mode Analysis}

\begin{table}[hbt!]
\caption{Vibrational properties of encounter complex (EC), first transition state (TS$_1$), transition complex (TC) and second transition state (TS$_2$) in respective principal axis frame of polarizability tensor obtained at CAM-B3LYP(D4)/def2-TZVPPD/CPCM(MeOH) level of theory with ORCA 5.0.4. Normal-mode frequencies, $\tilde{\nu}$, relative displacement of Si-C bond along $z$-axis, $\Delta r_z$, intensity, $I\propto\sum_\kappa T^2_\kappa$, frequency-weighted dipole derivatives, $T_\kappa=\frac{1}{\sqrt{2\omega_i}}\frac{\partial d_\kappa}{\partial Q_i}$, along Cartesian coordinates, $\kappa=x,y,z$, with $x$-axis perpendicular to the phenyl ring and $z$-axis parallel to the Si-C bond.}
\setlength\extrarowheight{3pt}
\begin{tabular}{ccccccccccc}
\hline\hline
mode & $\tilde{\nu}/\mathrm{cm}^{-1}$ & $\Delta r_z/\mathrm{a.u.}$ && $I/\mathrm{a.u.}$ && $T_x/\mathrm{a.u.}$ && $T_y/\mathrm{a.u.}$ &&  $T_z/\mathrm{a.u.}$ \\
\hline
$a^{\mathrm{EC}}_{38}$  &  862.1  & ---   && 0.01    && 0.09 && -0.05  &&  ---  \\
$a^{\mathrm{EC}}_{39}$  &  862.5  & ---   && 0.01    && 0.05 &&  0.09  &&  ---  \\
$a^{\mathrm{EC}}_{40}$  &  874.6  & 0.06  && 0.04    &&  ---  &&  ---   && -0.19 \\
\vspace{0.2cm}
\\
$a^{\mathrm{TS}_1}_{38}$  &  856.3  & 0.09   && 0.04    && ---    && ---   &&  0.19  \\
$a^{\mathrm{TS}_1}_{39}$  &  858.4  & 0.01   && 0.01    &&  0.1   && -0.03 &&  ---  \\
$a^{\mathrm{TS}_1}_{40}$  &  859.8  & 0.01   && 0.01    && -0.03  && -0.1  &&  --- \\
\vspace{0.2cm}
\\
$a^{\mathrm{TC}}_{38}$  &  838.1  &  ---   && 0.01    &&  0.08  && -0.05  &&   ---  \\
$a^{\mathrm{TC}}_{39}$  &  840.4  &  ---   && 0.01    &&  0.05  &&  0.08  &&   ---  \\
$a^{\mathrm{TC}}_{40}$  &  865.5  &  0.04  && 0.02    &&  ---   && -0.01  &&   0.16 \\
\vspace{0.2cm}
\\
$a^{\mathrm{TS}_2}_{38}$  &  857.1  & 0.01   && 0.016    && -0.09  && -0.08 && -0.04  \\
$a^{\mathrm{TS}_2}_{39}$  &  860.9  & 0.07   && 0.019    &&  0.02  && -0.06 &&  0.12  \\
$a^{\mathrm{TS}_2}_{40}$  &  861.6  & 0.05   && 0.016    &&  0.08  &&  0.07 &&  0.07 \\
\hline\hline
\end{tabular}
\label{tab.vib_details}
\end{table}

\begin{table}[hbt]
\caption{Selected normal modes of encounter complex (EC), first transition state (TS$_1$), transition complex (TC) and second transition state (TS$_2$) in respective principal axis frame of polarizability tensor obtained at CAM-B3LYP(D4)/def2-TZVPPD/CPCM(MeOH) level of theory with ORCA 5.0.4.}
\setlength\extrarowheight{3pt}
\begin{flushleft}
EC
\end{flushleft}
\begin{tabular}{ccccc}
\includegraphics[width=0.28\textwidth]{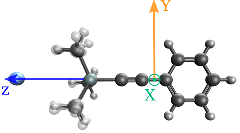} 
&&
\includegraphics[width=0.28\textwidth]{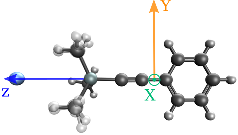}
&&
\includegraphics[width=0.28\textwidth]{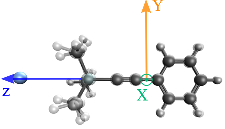}  
\vspace{0.2cm}
\\
$\hbar\omega_{a_{38}}=862.1\,\mathrm{cm}^{-1}$
&&
$\hbar\omega_{a_{39}}=862.5\,\mathrm{cm}^{-1}$
&&
$\hbar\omega_{a_{40}}=874.6\,\mathrm{cm}^{-1}$
\end{tabular}
\begin{flushleft}
TS$_1$
\end{flushleft}
\begin{tabular}{ccccc}
\includegraphics[width=0.28\textwidth]{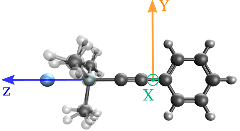} 
&&
\includegraphics[width=0.28\textwidth]{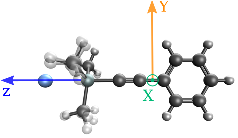}
&&
\includegraphics[width=0.28\textwidth]{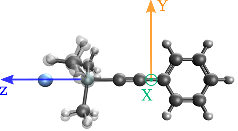}  
\vspace{0.2cm}
\\
$\hbar\omega_{a_{38}}=856.3\,\mathrm{cm}^{-1}$
&&
$\hbar\omega_{a_{39}}=858.4\,\mathrm{cm}^{-1}$
&&
$\hbar\omega_{a_{40}}=859.8\,\mathrm{cm}^{-1}$
\end{tabular}
\begin{flushleft}
TC
\end{flushleft}
\begin{tabular}{ccccc}
\includegraphics[width=0.28\textwidth]{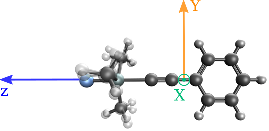} 
&&
\includegraphics[width=0.28\textwidth]{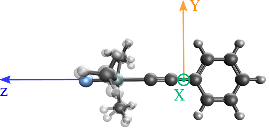}
&&
\includegraphics[width=0.28\textwidth]{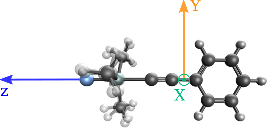}  
\vspace{0.2cm}
\\
$\hbar\omega_{a_{38}}=838.1\,\mathrm{cm}^{-1}$
&&
$\hbar\omega_{a_{39}}=840.4\,\mathrm{cm}^{-1}$
&&
$\hbar\omega_{a_{40}}=865.5 \,\mathrm{cm}^{-1}$
\end{tabular}
\begin{flushleft}
TS$_2$
\end{flushleft}
\begin{tabular}{ccccc}
\includegraphics[width=0.28\textwidth]{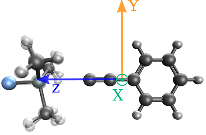} 
&&
\includegraphics[width=0.28\textwidth]{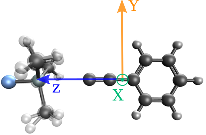}
&&
\includegraphics[width=0.28\textwidth]{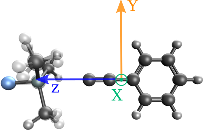}  
\vspace{0.2cm}
\\
$\hbar\omega_{a_{38}}=857.1\,\mathrm{cm}^{-1}$
&&
$\hbar\omega_{a_{39}}=860.9\,\mathrm{cm}^{-1}$
&&
$\hbar\omega_{a_{40}}=861.6\,\mathrm{cm}^{-1}$
\end{tabular}
\label{tab.vib_details_modes}
\end{table}

\newpage

\begin{table*}[hbt]
\caption{Structure of PTA in Cartesian coordinates (\AA) optimized on B3LYB(D4)/def2-TZVPPD/CPCM(MeOH) level of theory with body-fixed axis system aligned with principal axis system of polarizability tensor at CAM-B3LYB(D4)/def2-TZVPPD/CPCM(MeOH) level of theory. PTA forms reactants (R) with fluoride anion.}
\begin{tabular}{llll}
C  & -0.0000000000 & 0.0000000000  & 0.7123060606  \\
C  & -0.0002756537 & -0.0038831469 & 1.9264903604  \\
C  & 0.0000000000  & 0.0000000000  & -0.7123060606 \\
C  & -0.0022073230 & -0.0023224386 & -3.5085459475 \\
C  & 0.0000000000  & 1.2089194641  & -1.4241984560 \\
C  & -0.0008527785 & -1.2101268589 & -1.4222847164 \\
C  & -0.0018628984 & -1.2065472237 & -2.8098196293 \\
C  & -0.0011520961 & 1.2029817185  & -2.8118043863 \\
H  & 0.0002096550  & 2.1442840455  & -0.8814649612 \\
H  & -0.0003907425 & -2.1445385732 & -0.8779150763 \\
H  & -0.0019747200 & -2.1456345819 & -3.3472685792 \\
H  & -0.0017302601 & 2.1411471688  & -3.3508550554 \\
H  & -0.0031981930 & -0.0032456497 & -4.5905019281 \\
Si & -0.0052135629 & -0.0136557319 & 3.7693885065  \\
C  & 0.1946164394  & -1.7875382542 & 4.3395014123  \\
H  & 1.1397943108  & -2.2075902050 & 3.9903187209  \\
H  & -0.6166522216 & -2.4129494361 & 3.9622642930  \\
H  & 0.1830711241  & -1.8376009178 & 5.4308456812  \\
C  & 1.4216070487  & 1.0450269052  & 4.3655028778  \\
H  & 2.3764978325  & 0.6536281609  & 4.0097625233  \\
H  & 1.4473712655  & 1.0635911850  & 5.4576468276  \\
H  & 1.3208220788  & 2.0726699627  & 4.0111354953  \\
C  & -1.6389036244 & 0.6928002290  & 4.3567071798  \\
H  & -1.6759903452 & 0.6968593886  & 5.4487246826  \\
H  & -2.4781028744 & 0.0984881884  & 3.9905499784  \\
H  & -1.7695903823 & 1.7190942020  & 4.0085299456    
\end{tabular}
\end{table*}

\begin{table*}[hbt]
\caption{Structure of anionic encounter complex (EC) in Cartesian coordinates (\AA) optimized on B3LYB(D4)/def2-TZVPPD/CPCM(MeOH) level of theory with body-fixed axis system aligned with principal axis system of polarizability tensor at CAM-B3LYB(D4)/def2-TZVPPD/CPCM(MeOH) level of theory. }
\begin{tabular}{llll}
C  & -0.00442 & -0.00810 & -0.95144 \\
C  & -0.00477 & -0.00595 & 0.26275  \\
C  & -0.00442 & -0.00810 & -2.37601 \\
C  & -0.00528 & -0.00641 & -5.17250 \\
C  & -0.00442 & 1.20172  & -3.08642 \\
C  & -0.00476 & -1.21707 & -3.08780 \\
C  & -0.00513 & -1.21149 & -4.47539 \\
C  & -0.00483 & 1.19785  & -4.47397 \\
H  & -0.00469 & 2.13627  & -2.54230 \\
H  & -0.00431 & -2.15233 & -2.54488 \\
H  & -0.00502 & -2.14982 & -5.01418 \\
H  & -0.00535 & 2.13685  & -5.01158 \\
H  & -0.00574 & -0.00575 & -6.25446 \\
Si & -0.00212 & -0.00248 & 2.10700  \\
C  & 1.05794  & -1.43418 & 2.68743  \\
H  & 2.08287  & -1.33443 & 2.32505  \\
H  & 0.66041  & -2.38578 & 2.32937  \\
H  & 1.08387  & -1.46468 & 3.77899  \\
C  & 0.71110  & 1.63141  & 2.68318  \\
H  & 1.73350  & 1.76034  & 2.32300  \\
H  & 0.72649  & 1.67001  & 3.77466  \\
H  & 0.11297  & 2.46964  & 2.32079  \\
C  & -1.77089 & -0.20248 & 2.69357  \\
H  & -1.80678 & -0.20202 & 3.78539  \\
H  & -2.19685 & -1.14314 & 2.33923  \\
H  & -2.39805 & 0.61424  & 2.33131  \\
F  & 0.03284  & 0.05134  & 6.48236  
\end{tabular}
\end{table*}

\begin{table*}[hbt]
\caption{Structure of anionic first transition state (TS$_1$) in Cartesian coordinates (\AA) optimized on B3LYB(D4)/def2-TZVPPD/CPCM(MeOH) level of theory with body-fixed axis system aligned with principal axis system of polarizability tensor at CAM-B3LYB(D4)/def2-TZVPPD/CPCM(MeOH) level of theory. }
\begin{tabular}{llll}
C  & -0.00436 & 0.00941  & -0.73329 \\
C  & -0.00519 & 0.00842  & 0.48332  \\
C  & -0.00436 & 0.00941  & -2.15961 \\
C  & -0.00590 & 0.00839  & -4.96275 \\
C  & -0.00436 & 1.21676  & -2.87566 \\
C  & -0.00518 & -1.19849 & -2.87484 \\
C  & -0.00584 & -1.19517 & -4.26284 \\
C  & -0.00520 & 1.21243  & -4.26370 \\
H  & -0.00382 & 2.15268  & -2.33349 \\
H  & -0.00543 & -2.13401 & -2.33199 \\
H  & -0.00637 & -2.13468 & -4.80002 \\
H  & -0.00540 & 2.15154  & -4.80157 \\
H  & -0.00653 & 0.00800  & -6.04477 \\
Si & 0.00114  & -0.00346 & 2.38457  \\
C  & -0.07999 & -1.86136 & 2.66748  \\
H  & 0.79110  & -2.34099 & 2.21374  \\
H  & -0.96714 & -2.27206 & 2.17872  \\
H  & -0.10961 & -2.10273 & 3.72641  \\
C  & 1.65175  & 0.84878  & 2.68315  \\
H  & 2.45764  & 0.25512  & 2.24400  \\
H  & 1.84398  & 0.98168  & 3.74434  \\
H  & 1.66092  & 1.82359  & 2.18909  \\
C  & -1.56212 & 0.99483  & 2.69980  \\
H  & -1.75674 & 1.10417  & 3.76323  \\
H  & -2.41669 & 0.50474  & 2.22640  \\
H  & -1.46593 & 1.98558  & 2.24827  \\
F  & 0.01762  & -0.02830 & 4.88749  
\end{tabular}
\end{table*}

\begin{table*}[hbt]
\caption{Structure of anionic transition complex (TC) in Cartesian coordinates (\AA) optimized on B3LYB(D4)/def2-TZVPPD/CPCM(MeOH) level of theory with body-fixed axis system aligned with principal axis system of polarizability tensor at CAM-B3LYB(D4)/def2-TZVPPD/CPCM(MeOH) level of theory. }
\begin{tabular}{llll}
C  & -0.00775 & 0.01659  & -0.64497 \\
C  & -0.00743 & 0.01593  & 0.57723  \\
C  & -0.00775 & 0.01659  & -2.07217 \\
C  & -0.00914 & 0.01509  & -4.88465 \\
C  & -0.00775 & 1.22167  & -2.79499 \\
C  & -0.00848 & -1.18925 & -2.79382 \\
C  & -0.00910 & -1.18702 & -4.18205 \\
C  & -0.00850 & 1.21791  & -4.18326 \\
H  & -0.00753 & 2.15911  & -2.25497 \\
H  & -0.00834 & -2.12609 & -2.25276 \\
H  & -0.00936 & -2.12762 & -4.71789 \\
H  & -0.00876 & 2.15789  & -4.72017 \\
H  & -0.00971 & 0.01452  & -5.96670 \\
Si & 0.00571  & -0.01247 & 2.59835  \\
C  & -1.16456 & 1.49275  & 2.62589  \\
H  & -2.09885 & 1.24971  & 2.11224  \\
H  & -0.71439 & 2.32315  & 2.07540  \\
H  & -1.40467 & 1.83533  & 3.63303  \\
F  & 0.02421  & -0.05259 & 4.41818  \\
C  & -0.71395 & -1.77948 & 2.56141  \\
C  & 1.89372  & 0.26432  & 2.57772  \\
H  & 2.13272  & 1.17143  & 2.01640  \\
H  & 2.38600  & -0.56202 & 2.05744  \\
H  & 2.32717  & 0.35075  & 3.57474  \\
H  & -0.90766 & -2.18974 & 3.55312  \\
H  & -0.02347 & -2.44828 & 2.03991  \\
H  & -1.64674 & -1.79084 & 1.99159  
\end{tabular}
\end{table*}

\begin{table*}[hbt]
\caption{Structure of anionic second transition state (TS$_2$) in Cartesian coordinates (\AA) optimized on B3LYB(D4)/def2-TZVPPD/CPCM(MeOH) level of theory with body-fixed axis system aligned with principal axis system of polarizability tensor at CAM-B3LYB(D4)/def2-TZVPPD/CPCM(MeOH) level of theory. }
\begin{tabular}{llll}
C  & -0.18676 & 0.06101  & -1.01084 \\
C  & -0.18169 & 0.05732  & 0.22727  \\
C  & -0.18676 & 0.06101  & -2.43937 \\
C  & -0.18524 & 0.05962  & -5.26351 \\
C  & -0.18676 & 1.26411  & -3.17066 \\
C  & -0.18641 & -1.14282 & -3.16958 \\
C  & -0.18555 & -1.14122 & -4.55795 \\
C  & -0.18579 & 1.26111  & -4.55910 \\
H  & -0.18728 & 2.20322  & -2.63291 \\
H  & -0.18665 & -2.08149 & -2.63105 \\
H  & -0.18517 & -2.08292 & -5.09262 \\
H  & -0.18560 & 2.20225  & -5.09472 \\
H  & -0.18466 & 0.05907  & -6.34560 \\
Si & 0.17915  & -0.06003 & 3.19496  \\
C  & 1.72419  & 0.92161  & 2.80052  \\
H  & 1.68706  & 1.88893  & 3.30873  \\
H  & 1.83705  & 1.09025  & 1.73325  \\
H  & 2.60397  & 0.38797  & 3.17008  \\
C  & -1.45915 & 0.84314  & 3.04984  \\
H  & -2.19178 & 0.24227  & 2.51228  \\
H  & -1.33951 & 1.77735  & 2.50158  \\
H  & -1.85173 & 1.07073  & 4.04402  \\
C  & 0.22910  & -1.88375 & 2.75916  \\
H  & 0.60812  & -2.45826 & 3.60798  \\
H  & 0.86819  & -2.06369 & 1.89704  \\
H  & -0.76744 & -2.25399 & 2.51259  \\
F  & 0.32358  & -0.14536 & 4.86503 
\end{tabular}
\end{table*}

\begin{table*}[hbt]
\caption{Structure of anionic product complex (PC) in Cartesian coordinates (\AA) optimized on B3LYB(D4)/def2-TZVPPD/CPCM(MeOH) level of theory with body-fixed axis system aligned with principal axis system of polarizability tensor at CAM-B3LYB(D4)/def2-TZVPPD/CPCM(MeOH) level of theory. }
\begin{tabular}{llll}
C  & -0.00775 & 0.01659  & -0.64497 \\
C  & -0.00743 & 0.01593  & 0.57723  \\
C  & -0.00775 & 0.01659  & -2.07217 \\
C  & -0.00914 & 0.01509  & -4.88465 \\
C  & -0.00775 & 1.22167  & -2.79499 \\
C  & -0.00848 & -1.18925 & -2.79382 \\
C  & -0.00910 & -1.18702 & -4.18205 \\
C  & -0.00850 & 1.21791  & -4.18326 \\
H  & -0.00753 & 2.15911  & -2.25497 \\
H  & -0.00834 & -2.12609 & -2.25276 \\
H  & -0.00936 & -2.12762 & -4.71789 \\
H  & -0.00876 & 2.15789  & -4.72017 \\
H  & -0.00971 & 0.01452  & -5.96670 \\
Si & 0.00571  & -0.01247 & 2.59835  \\
C  & -1.16456 & 1.49275  & 2.62589  \\
H  & -2.09885 & 1.24971  & 2.11224  \\
H  & -0.71439 & 2.32315  & 2.07540  \\
H  & -1.40467 & 1.83533  & 3.63303  \\
F  & 0.02421  & -0.05259 & 4.41818  \\
C  & -0.71395 & -1.77948 & 2.56141  \\
C  & 1.89372  & 0.26432  & 2.57772  \\
H  & 2.13272  & 1.17143  & 2.01640  \\
H  & 2.38600  & -0.56202 & 2.05744  \\
H  & 2.32717  & 0.35075  & 3.57474  \\
H  & -0.90766 & -2.18974 & 3.55312  \\
H  & -0.02347 & -2.44828 & 2.03991  \\
H  & -1.64674 & -1.79084 & 1.99159  
\end{tabular}
\end{table*}

\begin{table*}[hbt]
\caption{Structure of PA anion (PA) in Cartesian coordinates (\AA) optimized on B3LYB(D4)/def2-TZVPPD/CPCM(MeOH) level of theory with body-fixed axis system aligned with principal axis system of polarizability tensor at CAM-B3LYB(D4)/def2-TZVPPD/CPCM(MeOH) level of theory. Forms products with FMS.}
\begin{tabular}{llll}
C & -0.00507 & -0.00032 & 3.34762  \\
C & 0.00147  & -0.00024 & 2.11037  \\
C & 0.00147  & -0.00024 & 0.68152  \\
C & 0.00147  & 1.20376  & -0.04755 \\
C & -0.00004 & 1.20164  & -1.43598 \\
C & -0.00038 & 0.00044  & -2.14058 \\
C & -0.00012 & -1.20114 & -1.43644 \\
C & 0.00117  & -1.20403 & -0.04807 \\
H & 0.00224  & 2.14226  & 0.49102  \\
H & -0.00121 & 2.14289  & -1.97123 \\
H & -0.00158 & 0.00030  & -3.22264 \\
H & -0.00113 & -2.14189 & -1.97257 \\
H & 0.00181  & -2.14278 & 0.49006 
\end{tabular}
\end{table*}

\begin{table*}[hbt]
\caption{Structure of FMS in Cartesian coordinates (\AA) optimized on B3LYB(D4)/def2-TZVPPD/CPCM(MeOH) level of theory with body-fixed axis system aligned with principal axis system of polarizability tensor at CAM-B3LYB(D4)/def2-TZVPPD/CPCM(MeOH) level of theory. Forms products with PA anion.}
\begin{tabular}{llll}
F  & 0.00008  & 0.00034  & 1.65234  \\
Si & 0.00008  & 0.00034  & 0.01414  \\
C  & 0.00008  & 1.78469  & -0.52118 \\
C  & -1.54603 & -0.89072 & -0.52153 \\
C  & 1.54601  & -0.89460 & -0.51420 \\
H  & 0.00311  & 1.84920  & -1.61189 \\
H  & 0.88339  & 2.30823  & -0.15007 \\
H  & -0.88768 & 2.30521  & -0.15618 \\
H  & -1.59857 & -0.92724 & -1.61230 \\
H  & -2.44078 & -0.38079 & -0.15865 \\
H  & -1.55967 & -1.91708 & -0.14930 \\
H  & 1.60778  & -0.92974 & -1.60441 \\
H  & 1.55031  & -1.92163 & -0.14327 \\
H  & 2.43978  & -0.39025 & -0.14148
\end{tabular}
\end{table*}


\end{document}